\documentclass[10pt,twocolumn,twoside,letterpaper]{IEEEtran}
\usepackage{geometry}

\usepackage{graphicx} 
\makeatletter
\def\ps@IEEEtitlepagestyle{
  \def\@oddfoot{\mycopyrightnotice}
  \def\@evenfoot{}
}
\def\mycopyrightnotice{
  {\footnotesize
  \begin{minipage}{\textwidth}
  \centering
  978-1-7281-4164-0/19/\$31.00 \copyright2019 IEEE
  \end{minipage}
  }
}
\usepackage{url}

\begin{document}
\title{Passive Gating Grid for Ion Back Flow Suppression in High Luminosity Collider Experiments}
%
%
%
\author{K. Dehmelt, P. Garg, T. K. Hemmick, A. Milov, E. Shulga and V. Zakharov 
\thanks{Manuscript submitted on December 12, 2019. Current work is supported in part by the US-Israel Binational Science Foundation grant number 2016240.}
\thanks{K. Dehmelt, P. Garg, T. K. Hemmick, and V. Zakharov are affiliated with Department of Physics and Astronomy, Stony Brook University, Stony Brook, USA.}
\thanks{A. Milov and E. Shulga are with the Department of Particle Physics, Weizmann Institute of Science, Rehovot, Israel.}}
\maketitle
\pagenumbering{gobble}

\begin{abstract}
Time Projection Chamber (TPC) is one of the main tracking systems for many current and future collider experiments at RHIC and LHC. It has a capability to measure the space points of charged tracks for good momentum resolution as well as the energy loss (dE/dx) for particle identification with good energy resolution. Both of these features depend strongly on the amount of space charge in the TPC gas volume, mainly due to the ions from the amplification stage.  An active gating grid has been used thus far to gate the electrons and ions by switching the polarities of the grid wires. Therefore, active gating does introduce a limitation for data taking rates in high luminosity collisions. In this work we propose several options of a passive gating, where a significant reduction of Ion Back Flow (IBF) is possible in a high luminosity environment without any dead time issues due to gating operation. Particularly, the application of a TPC passive gating for the sPHENIX experiment at RHIC is presented, which is currently under development.
\end{abstract}


\section{Introduction}
%
%
%
%
\IEEEPARstart{A}{}Time Projection Chamber is one of the main tracking systems for the proposed sPHENIX experiment at RHIC to resolve the Upsilon states via their di-electron channel. The strong magnetic field of the BaBar solenoid, a Neon based fast gas mixture, and a high drift electric field will improve the E-field distortions due to ion backflow significantly in the current sPHENIX TPC design. Furthermore, a quadruple GEM stack arranged in a particular orientation and hole pattern \cite{ALICE:2014qrd} or a MicroMegas based amplification \cite{Colas:2004tg} will also help to reduce IBF. Nonetheless, due to the expected high luminosity at sPHENIX ($\sim$15kHz), including a passive gating grid can be helpful to avoid the space charge distortions.
We have studied several options to achieve good electron transparency ($>$80\%)for the primary electrons and high blocking for the ions coming from the amplification stage. We have simulated woven wire meshes, different patterns of etched meshes, hexagonal micro-pattern meshes and static bi-polar wire gating. 

Typically, the readout chambers are operated with an active bipolar Gating Grid (GG) in the presence of a trigger, which switches to the transparent mode to allow the ionization electrons to pass into the amplification region. After the maximum drift time for electrons the GG wires are biased with an alternating voltage, which makes the grid opaque to electrons and ions. This protects the amplification region against unwanted ionization from the drift region, and prevents back-drifting ions to enter the drift volume. In particular, the latter would lead to significant space-charge distortions. Due to the low mobility of ions, efficient ion blocking requires the GG to remain closed after the end of the event readout, corresponding to the typical time it takes the ions to drift from the amplification to the GG. This gating scheme leads to an intrinsic dead time for the TPC system, implying a principal rate limitation of the TPC. 

A concept of static bi-polar passive gating is also explored in the present work, where the alternate polarity of wires is used to block the ions with good primary electron transparency at the same time, in the presence of strong magnetic field. The magnetic field helps to pass the electrons up to the optical transparency of the grid, while ions are completely blocked.
Furthermore, we propose and briefly discuss a modified Zigzag pad shape to compensate the distortion in an electron's position due to GG implementation.

\section{RESULTS AND DISCUSSION}

\begin{figure}[htbp]
\begin{center}
\includegraphics[scale=0.41]{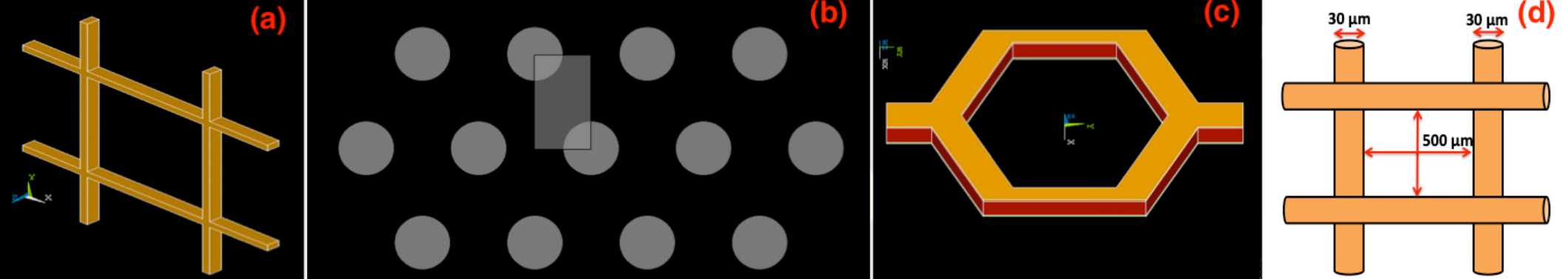}
\caption{Different configurations for static bi-polar mesh.}
\label{Fig1}
\end{center}
\end{figure}

In this section, we show the results for various options for passive gating, which can be coupled with TPC for space charge reduction. We have done simulation studies using ANSYS, a Finite Element Method (FEM) based software \cite{ansys} and CERN based Garfield++ simulation package~\cite{garfield}. Fig.\ref{Fig1} shows various shapes of elements like (a) etched square mesh, (b) cylindrical hole patterns, (c) hexagonal micro-pattern mesh and (d) a simple wire mesh. 

The electric field maps are generated with ANSYS and are transported in Garfield++ to simulate the electrons and ions in the presence of different gas mixtures and magnetic field along the drift field. The electron transparency corresponds to the fraction of electrons passing through the gating, while ion blocking corresponds to the fraction of ions collected on the gating grid or blocked from entering the drift volume. The gas attachment properties are also included in these simulations. 

\begin{figure}[htbp]
\begin{center}
\includegraphics[scale=0.32]{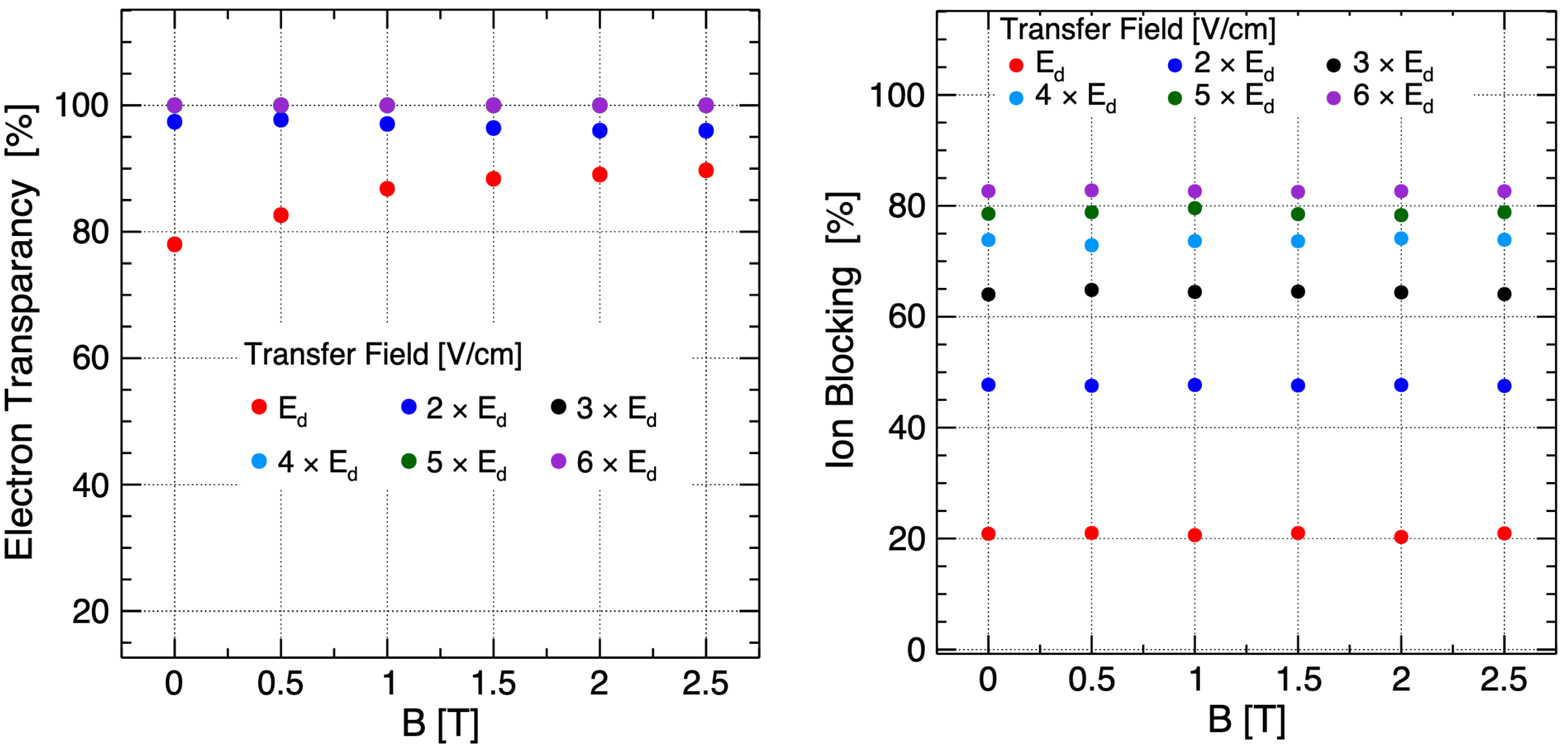}
\caption{(left) Electron Transparency  and (right) Ion Blocking are shown as a function of magnetic field for a wire mesh gating structure. Legends show the  strength of transfer fields for a fix drift field of 400 V/cm in Ne+CF$_{4}$+iC$_{4}$H$_{10}$(95:3:2) gas mixture.}
\label{Fig2}
\end{center}
\end{figure}

\begin{figure}[htbp]
\begin{center}
\includegraphics[scale=0.32]{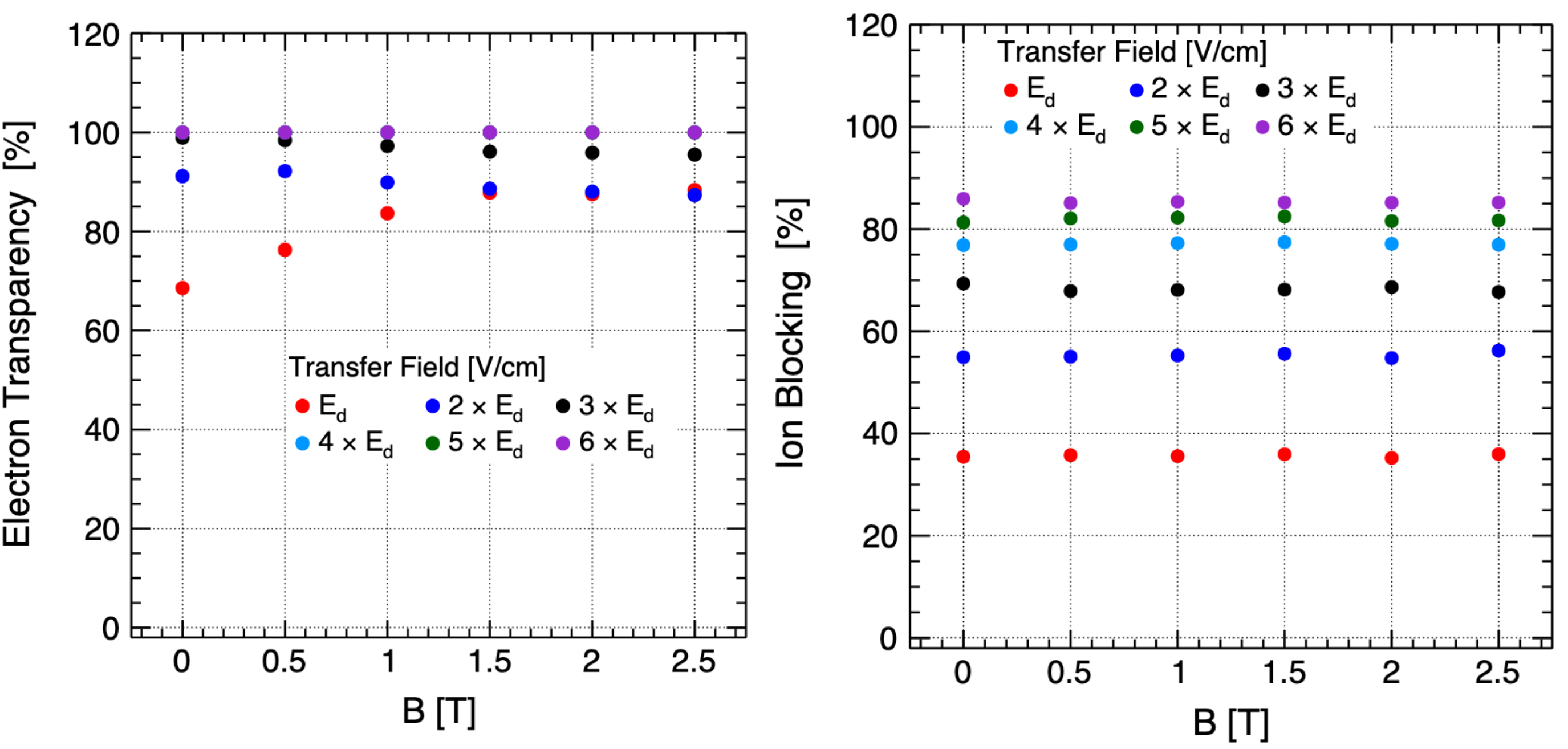}
\caption{(left) Electron Transparency  and (right) Ion Blocking are shown as a function of magnetic field for an etched rectangular mesh structure. Legends show the  strength of transfer fields for a fix drift field  of 400 V/cm in Ne+CF$_{4}$+iC$_{4}$H$_{10}$(95:3:2) gas mixture.}
\label{Fig3}
\end{center}
\end{figure}

Figs. \ref{Fig2} and \ref{Fig3} show the results for wire mesh (Fig.\ref{Fig1}d) and etched mesh (Fig.\ref{Fig1}a) for one particular configuration.
It is seen in Figs. \ref{Fig2} and \ref{Fig3}, that a significant amount of ion reduction and electron transparency can be achieved  by increasing the ratio between the transfer to the drift electric field. Increase in this ratio causes more ion drift lines to terminate at the gating structure, which helps to reduce the ion back flow. 

We have observed in our simulation studies that by simply putting a passive gating and tuning the field ratios, the ion blocking of $\sim$80\% and a corresponding electron transparency of $\sim$ 95\% can be achieved.  

\begin{figure}[htbp]
\begin{center}
\includegraphics[scale=0.8]{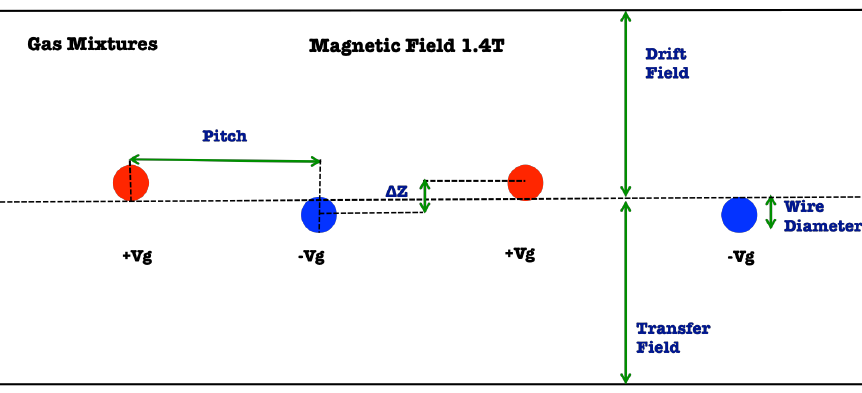}
\caption{Schematic of parameter space for wire gating.}
\label{Fig4}
\end{center}
\end{figure}

A bi-polar static gating grid as an ion blocking device is also explored.  A large parameter space is studied to find a suitable configuration for TPC gating, as is shown in Fig.\ref{Fig4}. 

\begin{figure}[htbp]
\begin{center}
\includegraphics[scale=0.33]{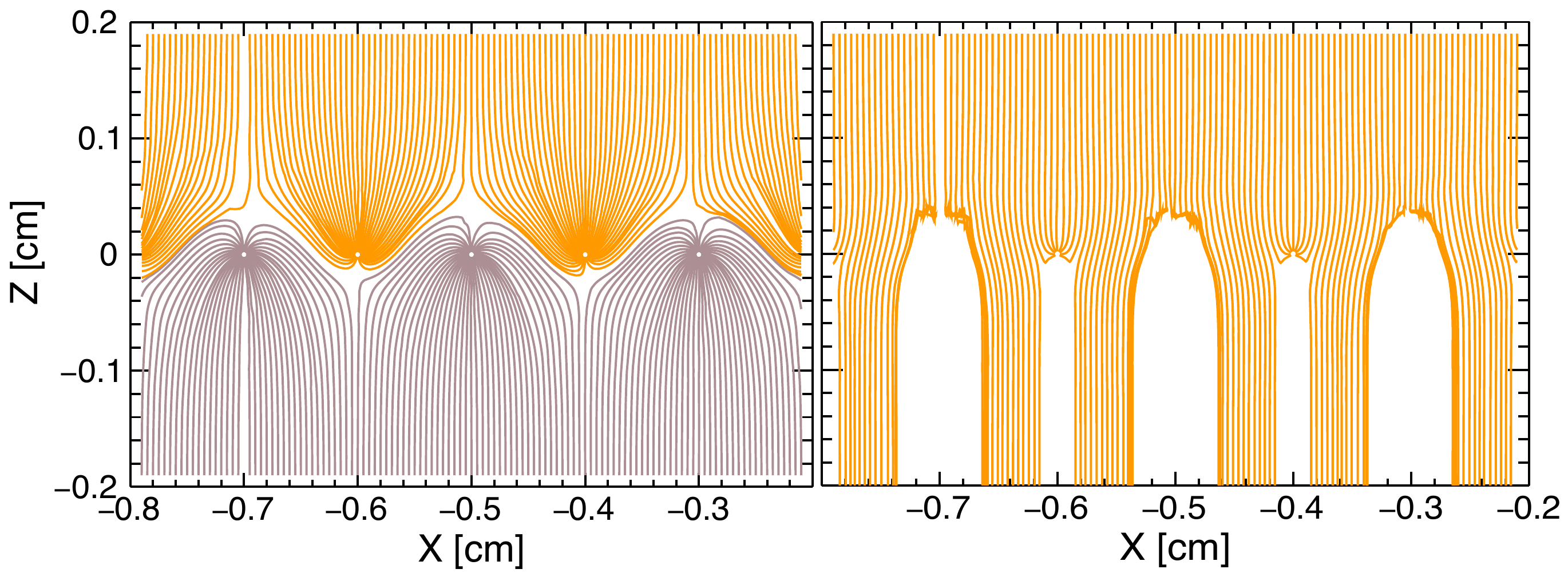}
\caption{(left) Yellow lines represent the electrons drift lines and brown lines show the ion drift lines in the absence of magnetic field.  (right) Electron drift lines when the magnetic field is turned on.}
\label{Fig5}
\end{center}
\end{figure}

The electron and ion trajectories (yellow and brown lines are for electrons for ions respectively) without magnetic field are shown in the left panel of Fig.~\ref{Fig5}. Once the magnetic field is turned on, the electron's trajectories are shown in the right panel of Fig.~\ref{Fig5}, although the ions remain unchanged. The cause for the electron transparency in the  presence of magnetic field can be understood as follows.

The motion of charged particles under the influence of electric and magnetic field (E and B) is governed by an equation of motion called Langevin Equation:

\begin{equation}
m \frac{du}{dt} = eE + e[u\times B] - Ku,
\end{equation}

where m and e are the mass and electric charge of the particle, u is its velocity, and K describes a frictional force proportional to u that is caused by the interaction of the particle with the gas. As is shown in the left panel of Fig. \ref{Fig5}, the electrons and ions are completely blocked, as they land on the wires depending on their polarities. But, as soon as the magnetic field is turned on the u$\times$B term of Langevin equation forces the electrons to go away from the wire, as is shown in the right panel of Fig.\ref{Fig5}.

\begin{figure}[htbp]
\begin{center}
\includegraphics[scale=0.28]{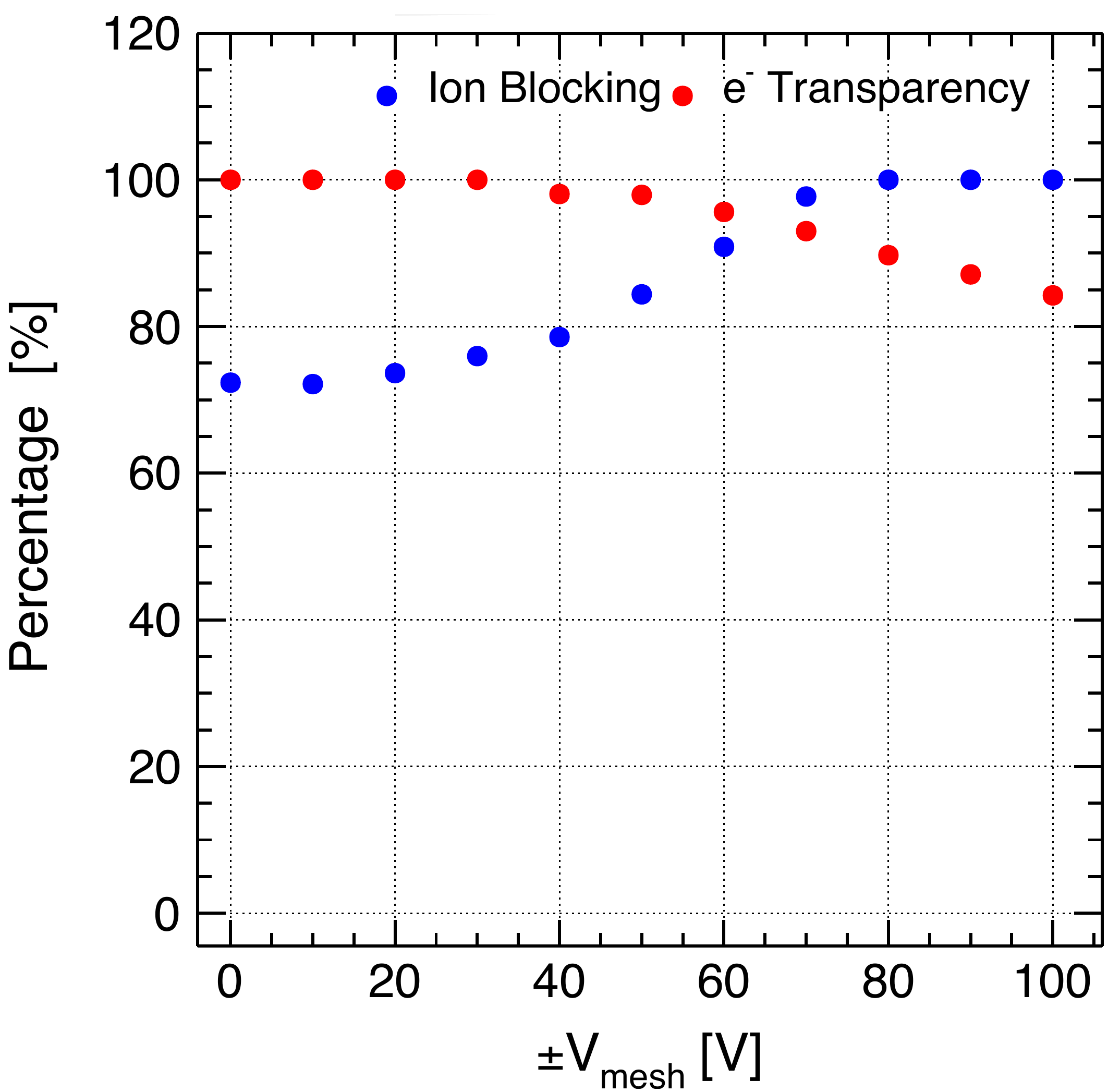}
\caption{ Simulation results in percentage as a function of potentials at the wire, with Ne:CF$_4$ (90:10) gas, 200V/cm drift field, 800V/cm transfer field, 90$\mu$m wire diameter and 1.5mm wire pitch.}
\label{Fig6}
\end{center}
\end{figure}

We have explored a huge parameter space for  wire gating, such as wire diameter, pitch, potential differences and gas mixtures for different magnetic fields as is shown in Fig.\ref{Fig4}. 
As an example, the percentage of electron transparency and ion blocking is shown in Fig.~\ref{Fig6} for one such combination of parameters.

Hence, tuning the wire configuration and selectively applying the potential on wires can lead to an optimized ion blocking and electron transparency.

As is evident from Fig.\ref{Fig5}, introducing a wire grid for gating also introduces distortions in the primary electron's trajectories, which can deteriorate the position resolution for TPC tracking. Further, to correct the distortions in electron's position due to gating we have explored several readout pad shapes, where these distortions can be compensated for high resolution TPC tracking.

\begin{figure}[htbp]
\begin{center}
\includegraphics[scale=0.33]{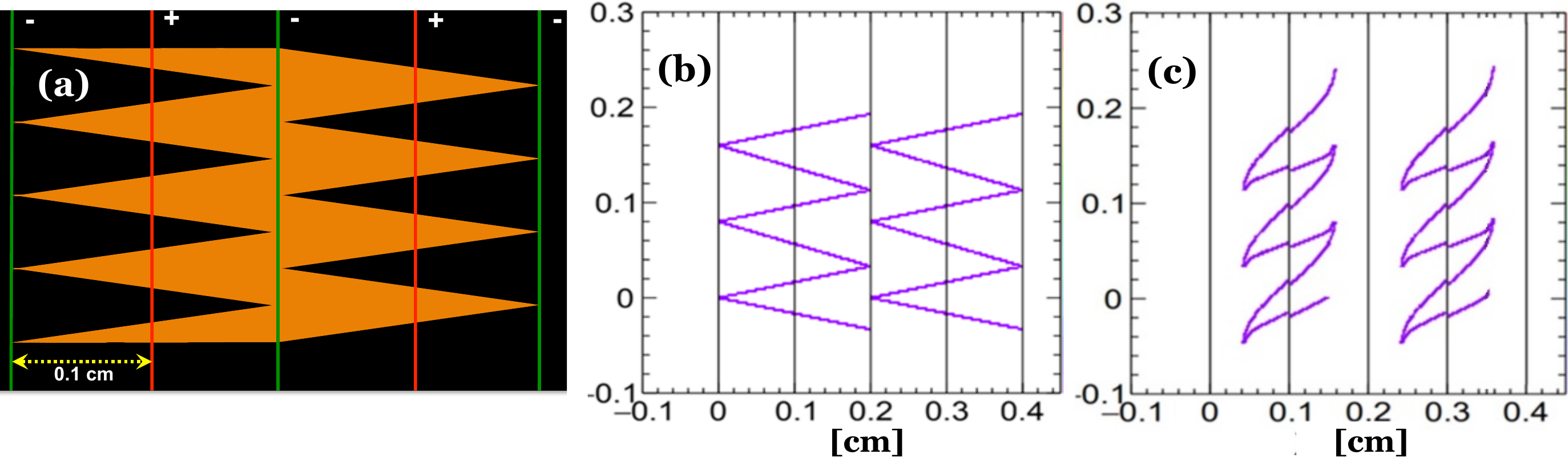}
\caption{(a) One possible wire-to-pad configuration. Simulations of (b) initial and (c) final $e^-$ positions propagated through the wires, without any diffusion, showing original and distorted pad edges.}
\label{Fig7}
\end{center}
\end{figure}

The readout pad plane design of sPHENIX TPC exploits the fact that by maximizing the charge sharing between neighboring ZigZag pads~\cite{Azmoun:2018ail}, a  high position resolution can be achieved.  Fig.\ref{Fig7}(a) shows a typical ZigZag pad sector view from top where the negative (green lines) and positive (red lines) potential wires are arranged in a particular fashion to serve as a GG. Fig.\ref{Fig7}(b) shows that the electron's initial position is chosen in such a way that they are focused along the edges of a ZigZag pad and the vertical lines are shown for eye guidance of the wire's position. The electrons are then transported through the gating grid while keeping the diffusion off such that the their position after passing through the grid is purely influenced by the electric field distortions.
Fig.\ref{Fig7}(c) shows the end coordinates of the electrons after going through the wire grid. 
 
Therefore, if one starts with a modified pad shape similar to Fig\ref{Fig7}(c), all the initial electrons can still be collected on the same pad despite facing the distortions, while conserving the position resolution. More simulation studies are underway to better understand the modified shapes of pad readouts,  influence of different gating configurations and gas properties etc.

\section{SUMMARY}
TPCs in high luminosity collider experiments require a passive gating option to cope with the high event rates. In the present work we demonstrate various options for the passive gating configurations, where a good ion blocking ($>$80\%) and electron transparencies ($>$90\%) can be achieved. Further reduction in IBF after adding a gating grid in addition to an already low IBF Quad-GEM configuration or in a MicroMegas configuration will be helpful to avoid space-charge distortions in the TPC.  The preparations are ongoing to measure these simulated results with real detector setup to establish the proof of principle.


\end{document}